\documentclass[12pt]{article}
\makeatletter
\def\nothing#1{}
\newdimen\earraycolsep
\renewcommand{\thetable}{\arabic{table}}
\renewcommand{\thefigure}{\arabic{figure}}
\renewcommand{\title}[1]{%
  \vspace*{80\p@}%
  {\parindent \z@ \raggedright \reset@font
    \bfseries #1\par
    \nobreak
    \vskip 36\p@
  }}
\def\author#1{{\pretolerance=10000 \raggedright \advance \leftskip by 1in \noindent #1 \vskip 1pc}}
\def\affiliation#1{{\advance\leftskip by 1in \noindent #1 \vskip -1pc}}

\renewcommand\section{\@startsection{section}{1}{\z@}{2pc \@plus 
      1ex minus .2ex}{1pc \@plus .2ex}{\reset@font
      \normalsize\bfseries\noindent
      {\addtocounter{section}{1}}\arabic{section}\ 
      {\setcounter{subsection}{0}
      \setcounter{subsubsection}{0}\setcounter{equation}{0}} }}
\renewcommand\subsection{\@startsection{subsection}{2}{\z@}{1pc \@plus 1ex
    minus.2ex}{1pc \@plus .2ex}
    {\reset@font\normalsize\bfseries
    \noindent{\addtocounter{subsection}{1}}%
    {\setcounter{subsubsection}{0}}\arabic{section}.\arabic{subsection}\ }}
\renewcommand\subsubsection{\@startsection{subsubsection}{3}{\parindent}
        {1pc \@plus 1ex minus.2ex}{-0.5em}{\reset@font\normalsize\bfseries%
        {\addtocounter{subsubsection}{1}} \hspace*{.6cm}
        \arabic{section}.\arabic{subsection}.\arabic{subsubsection}
        \hspace*{-7mm}}}

\def\AmS{{\protect\the\textfont2%
        A\kern-.1667em\lower.5ex\hbox{M}\kern-.125emS}}

\def\p@LaTeX{{\family{times}\series{m}\shape{n}\selectfont L\kern-.36em\raise.3ex\hbox{\scriptsize A}\kern-.15em T\kern-.1667em\lower.7ex\hbox{E}\kern-.125emX}}

\newlength{\colwidth}

\def\@oddhead{\hfil}
\def\@evenhead{\hfil}
\def\@oddfoot{{\bfseries\hfil\thepage}}
\def\@evenfoot{{\bfseries\thepage\hfil}}
\def\fnum@figure{\footnotesize\raggedright{\bfseries \figurename~\thefigure.}}
\def\fnum@table{\normalsize\raggedright{\bfseries \tablename~\thetable.}}
\long\def\@makecaption#1#2{\vskip 10\p@ {#1 #2\par}}
\long\def\@makefntext#1{\setbox0=\hbox{$\m@th^{\@thefnmark}$}\noindent\hangindent=\wd0 \box0 #1}
\def\centerfig#1#2#3#4{\vspace*{#2}\relax\centerline{\hbox to#1{\special{#4:#3.#4 x=#1, y=#2}\hfil}}}
\newbox\@atbox
\long\def\atable#1#2#3{\begin{table}[tbp]\centering\footnotesize
\setbox\@atbox\hbox{#2}
\parbox{\wd\@atbox}{\caption{#1}}\par\smallskip
#2
\par\smallskip\parbox{\wd\@atbox}{\raggedright #3}
\end{table}}
\def\@nbibitem#1{\noindent \hangindent=2pc \hangafter=1
\refstepcounter{enumi}\hbox to 2pc{\arabic{enumi}.\hfil}%
\immediate\write\@auxout{\string\bibcite{#1}{\arabic{enumi}}}}
\def\numbibliography{%
\section*{REFERENCES}%
\bgroup\footnotesize
\setcounter{enumi}{0}%
\def\newblock{\hskip .11em plus.33em minus.07em}%
\let\bibitem\@nbibitem}
\def\endnumbibliography{\par\egroup}
\makeatother

\usepackage{amssymb}

\def\1#1{{\bf #1}}
\def\2#1{{\cal #1}}
\def\3#1{{\sl #1}}
\def\4#1{{\tt #1}}
\def\5#1{{\sf #1}}
\def\6#1{{\mathfrak #1}}
\def\7#1{{\mathbb #1}}

\newcommand{\be}{\begin{equation}}
\newcommand{\ee}{\end{equation}}
\newcommand{\ba}{\begin{array}}
\newcommand{\ea}{\end{array}}
\newcommand{\bea}{\begin{eqnarray}}
\newcommand{\eea}{\end{eqnarray}}
\newcommand{\bean}{\begin{eqnarray*}}
\newcommand{\eean}{\end{eqnarray*}}
\newcommand{\nn}{\nonumber}

\newcommand{\ve}{\varepsilon}

\newcommand{\impl}{\Rightarrow}
\newcommand{\restr}{\upharpoonright}
\newcommand{\ol}{\overline}

\newcommand{\qft}{quantum field theory}
\newcommand{\qfts}{quantum field theories}

\newcommand{\npb}{Nucl. Phys. \1B}
\newcommand{\cmp}{Commun. Math. Phys. }

\newcommand{\rmp}{Rev. Math. Phys. }

\newtheorem{defin}{Definition}[section]
\newtheorem{lemma}[defin]{Lemma}
\newtheorem{prop}[defin]{Proposition}
\newtheorem{theorem}[defin]{Theorem}
\newtheorem{corr}[defin]{Corollary}

\newcommand{\bdefin}{\begin{defin}}
\newcommand{\blemma}{\begin{lemma}}
\newcommand{\bprop}{\begin{prop}}
\newcommand{\btheor}{\begin{theorem}}
\newcommand{\bcorr}{\begin{corr}}
\newcommand{\edefin}{\end{defin}}
\newcommand{\elemma}{\end{lemma}}
\newcommand{\eprop}{\end{prop}}
\newcommand{\etheor}{\end{theorem}}
\newcommand{\ecorr}{\end{corr}}

\begin{document}\noindent
{\sf DESY 96-237} \hfill {\sf ISSN 0418-9833} \\
{\sf November 1996} \hfill \\
\title{\\DISORDER OPERATORS, QUANTUM DOUBLES, AND HAAG DUALITY IN $1+1$ DIMENSIONS\footnote{Seminar given at the NATO Advanced Study Institute on {\it Quantum Fields and Quantum Spacetime}, Carg\`{e}se, 1996. To appear in the proceedings.}}

\author{Michael M\"uger}

\affiliation{II.\ Institut f\"ur Theoretische Physik, Universit\"at Hamburg\\
Luruper Chaussee 149, D--22761~Hamburg, Germany\\
Email: mueger@x4u2.desy.de}

\section{INTRODUCTION AND PREREQUISITES}        

Since the notion of the `quantum double' was coined by Drinfel'd in his famous
ICM lecture \cite{drin1} there have been several attempts aimed at a clarification
of its relevance to two dimensional \qft. The quantum double appears implicitly
in the work \cite{dvvv} on orbifold constructions in conformal field theory, where
conformal \qfts\ (CQFTs) are considered whose operators are fixpoints under the 
action of a symmetry group on another CQFT. Whereas the
authors emphasize that `the fusion algebra of the holomorphic G-orbifold
theory naturally combines both the representation and class algebra of the group G'
the relevance of the double is fully recognized only in \cite{dpr}. 
The quantum double also appears in the context of integrable \qfts, e.g.\
\cite{bern}, as well as in certain lattice models (e.g.\ \cite{szlvec}).
Common to these works is the role of disorder operators or `twist fields' which are 
`local with respect to $\2A$ up to the action of an element $g\in G$' \cite{dvvv}.

In this note, which is a compressed version of \cite{mue1}, we will use the methods of 
algebraic \qft\ \cite{haag,K} to demonstrate the role of the quantum double as a hidden
symmetry in {\it every} \qft\ with group symmetry in $1+1$ dimensions fulfilling 
(besides the usual assumptions like locality) only two technical assumptions (Haag 
duality and split property, see below) but independent of conformal covariance or 
exact integrability.
As in \cite{dhr1} we will consider a \qft\ to be specified by a net of von Neumann 
algebras, i.e.\ a map $\2O\mapsto\2F(\2O)$
which assigns to any bounded region in $1+1$ dimensional Minkowski space a von 
Neumann algebra (i.e.\ an algebra of bounded operators closed under hermitian 
conjugation and weak limits) on the common Hilbert space $\2H$ such that
isotony holds:
\be \2O_1\subset\2O_2 \ \impl \ \2F(\2O_1)\subset\2F(\2O_2) .\ee
The quasilocal algebra $\2F=\ol{\bigcup_{\2O\in\2K}\2F(\2O)}^{\|\cdot\|}$, $\2K$ being 
the set of all double cones (intersections of forward and backward lightcones), is 
assumed to be irreducible: $\2F'=\7C\11$.\footnote{In general $\2M'=\{X\in\2B(\2H)|XY=YX\,\forall Y\in\2M\}$ denotes the algebra of all bounded operators commuting with all operators in $\2M$.}
In order to simplify the exposition we restrict ourselves in this note to pure Bose
fields (for the case of general Bose-Fermi commutation relations see \cite{mue1}):
\be \2F(\2O)\subset\2F(\2O')' .\label{commrel}\ee

Poincar\'{e} covariance is implemented by assuming the existence of a (strong\-ly
continuous) unitary representation on $\2H$ of the Poincar\'{e} group $\2P$ such that
\be \alpha_{(\Lambda,a)}(\2F(\2O))=Ad\, U(\Lambda,a)(\2F(\2O))=
   \2F(\Lambda\2O+a) .\ee
The spectrum of the generators of the translations (momenta) is required to be 
contained in the closed forward lightcone and the existence of a unique vacuum vector
$\Omega$ invariant under $\2P$ is assumed. Covariance under the conformal 
group, however, is {\it not} required.

Our last postulate (for the moment) concerns the inner symmetries of the 
theory. There shall be a compact group $G$, represented in a strongly
continuous fashion by unitary operators on $\2H$ leaving invariant the vacuum
such that the automorphisms $\alpha_g(F)=Ad\, U(g)(F)$ of $\2B(\2H)$ 
respect the local structure:
\be \alpha_g(\2F(\2O))=\2F(\2O) .\ee
The action may be assumed faithful, i.e.\ $\alpha_g\ne\mbox{id}\ \forall g\ne e$. 
(Compactness of $G$ need in fact not be postulated, as it is known to follow from 
the split property which will be introduced later. For the sake of simplicity we assume
in this note that the group $G$ commutes with the Poincar\'{e} group, see \cite{mue1}
and \cite[Appendix]{mue2} for further discussion.)
The observables are now defined as the gauge invariant operators:
\be \2A(\2O) = \2F(\2O)^G = \2F(\2O)\cap U(G)' .\label{avono}\ee

This framework was the starting point for the investigations in \cite{dhr1}
where in particular properties of the observable net (\ref{avono}) and its
representations on the sectors in $\2H$, i.e.\ the G-invariant subspaces, were 
studied. One important notion examined in \cite{dhr1} was
that of {\it duality} designating a maximality property in the sense that 
the local algebras cannot be enlarged (on the same Hilbert space) without violating 
spacelike commutativity. The postulate of duality for the fields consists 
in strengthening the locality postulate (\ref{commrel}) to
\be \2F(\2O) = \2F(\2O')'   \label{twduality} ,\ee
which means that $\2F(\2O')$, the von Neumann algebra generated by all 
$\2F(\2O_1)$, $\2O'\supset\2O_1\in\2K$ contains all operators commuting
with $\2F(\2O)$. (This can easily be generalized to the case with fermions.) Duality
has been proved to hold for free massive and massless scalar and Dirac fields in all 
dimensions as well as for several interacting theories ($P(\phi)_2, Y_2$).

From this it has been derived \cite[Theorem 4.1]{dhr1} (for $\ge 2+1$ dimensions) that 
duality holds for the observables when restricted to a simple sector $\2H_1$:
\be \6A(\2O)\equiv\2A(\2O)\restr\2H_1\ \Longrightarrow\ \6A(\2O)'=\6A(\2O')\ \   \forall\2O
\label{duality} .\ee
A sector $\2H_1$ is called simple if the group $G$ acts on it via multiplication with
a character
\be U(g)\restr\2H_1  = \chi(g)\cdot\11\restr\2H_1 .\ee
Clearly the vacuum sector is simple.
Furthermore it has been shown \cite[Theorem 6.1]{dhr1} that the irreducible 
representations of the observables on the charge sectors in $\2H$ are strongly 
locally equivalent to the vacuum representation in the sense that for any representation
$\pi(A)=A\restr\2H_\pi$ and any $\2O\in\2K$ 
\be \pi\restr\2A(\2O')\cong\pi_0\restr\2A(\2O') .\label{dhr}\ee

The fundamental facts (\ref{duality}) and (\ref{dhr}), which have come to be called
Haag duality and the DHR criterion respectively, were taken as starting points in
\cite{dhr3} where a more ambitious approach to the theory of superselection
sectors was advocated and developed to a large extent. The basic idea was that the
physical content of any \qft\ should reside in the observables and their vacuum
representation whereas all other physically relevant representations as well as 
unobservable charged fields interpolating between those and the vacuum sector should be 
constructed from the observable data. The vacuum representation was postulated
to satisfy (\ref{duality}), while (\ref{dhr}) was chosen as a selection criterion for
a class of interesting representations. It may be considered as one of the
triumphs of the algebraic approach that it has finally been possible to prove
\cite[and references given there]{dr2} the existence of a compact group $G$ `describing'
the structure of the DHR sectors and of an essentially unique net of field algebras 
acted upon by $G$ and generating the charged sectors from the vacuum.

In 1+1 dimensions part of the analysis breaks down due to the topological 
pecularity that the spacelike complement of a bounded (connected) region consists 
of two connected components as a consequence of which the permutation group governing 
the statistics is replaced by the braid group. The algebraic formalism of
\cite{dhr3} was adapted to this situation in \cite{frs}, see also \cite{mue2}.
It is still not known by which structure the compact group appearing
in the higher dimensional situation has to be replaced if a completely general 
solution to this question exists at all. 
Even though in $1+1$ dimensions one can not conclude the existence of a field net with
group symmetry it appears interesting to study nets of observables arising as fixpoint
nets (`orbifold theories'). This is the aim of the research to be reported here which
in particular leads to a complete understanding of another peculiarity in $1+1$ 
dimensions as will be discussed in the last section.

\section{SPLIT PROPERTY, DISORDER OPERATORS, AND NONLOCAL FIELD EXTENSIONS}
We begin by introducing some notation.
For any double cone $\2O\in\2K$ we designate the left and right spacelike 
complement by $W^\2O_{LL}$ and $W^\2O_{RR}$, respectively. Furthermore we write
$W^\2O_L$ and $W^\2O_R$ for ${W^\2O_{RR}}'$ and ${W^\2O_{LL}}'$. These
regions are wedge shaped, i.e.\ translates of the standard wedges
$W_L=\{x\in\7R^2 \mid x^1<-|x^0|\}$ and $W_R=\{x\in\7R^2 \mid x^1>|x^0|\}$.
With these definitions we have
$\2O=W^\2O_L \cap W^\2O_R$ and $\2O'=W^\2O_{LL} \cup W^\2O_{RR}$ which
graphically looks as in Figure 1.
\begin{figure}
\[\ba{c}
\begin{picture}(300,100)(-150,-50)\thicklines
\put(20,0){\line(-1,1){40}}
\put(20,0){\line(-1,-1){40}}
\put(-20,0){\line(1,1){40}}
\put(-20,0){\line(1,-1){40}}
\put(20,0){\line(1,1){40}}
\put(20,0){\line(1,-1){40}}
\put(-20,0){\line(-1,1){40}}
\put(-20,0){\line(-1,-1){40}}
\put(-5,-5){$\2O$}
\put(-70,-5){$W^\2O_{LL}$}
\put(45,-5){$W^\2O_{RR}$}
\put(-30,15){$W^\2O_L$}
\put(10,15){$W^\2O_R$}
\end{picture}
\ea\]
\caption{Wedges associated to a double cone}
\end{figure}
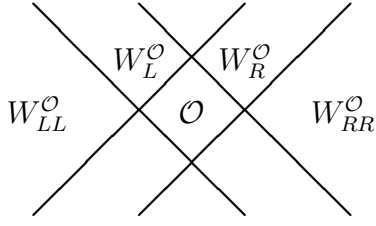
In analogy to ideas in statistical mechanics we introduce the notion of a family of
{\it disorder operators} which consists, for any $\2O\in\2K$ and any $g\in G$, of two 
unitary operators $U^\2O_L(g)$ and $U^\2O_R(g)$ verifying 
\be\ba{ccccc} \mbox{Ad}\,U_L^\2O(g)\restr\2F(W^\2O_{LL}) &=& \alpha_g &=&
   \mbox{Ad}\,U_R^\2O(g)\restr\2F(W^\2O_{RR}) ,\\
   \mbox{Ad}\,U_L^\2O(g)\restr\2F(W^\2O_{RR}) &=& \mbox{id} &=&
   \mbox{Ad}\,U_R^\2O(g)\restr\2F(W^\2O_{LL})  .
\ea\label{disord}\ee
A disorder operator thus interpolates between the action of an unbroken global symmetry
on one wedge and the trivial action on a wedge properly contained in the spacelike 
complement of the first one. 

In general it is not obvious that disorder operators exist. Therefore we introduce as
another axiom the {\it split property for wedges} which formalizes a strong form of
statistical independence of spacelike separated regions. A net of field algebras has
this property if for every double cone $\2O$ the von Neumann algebra 
$\2F(W^\2O_{LL})\vee\2F(W^\2O_{RR})$ is algebraically isomorphic to the tensor product
$\2F(W^\2O_{LL})\otimes\2F(W^\2O_{RR})$. It is known that free massive scalar and Dirac
fields satisfy this property and it seems reasonable to expect this to be the case in
every well behaved massive theory. For a discussion of related properties and for further
references we refer to the detailed review \cite{sum}.

Using essentially the same methods as in \cite{bdl} one can show that the split property
implies the existence of disorder operators $U^\2O_L(g),\,U^\2O_R(g)$ for all 
$\2O\in\2K,\ g\in G$. Besides the above defining equations these operators have the
following additional properties:
\be \left[  U^{\2O}_L (g), U^{\2O}_R (h) \right] = 0 ,\ \ \
   U^{\2O}_L (g) \ U^{\2O}_R (g) = U(g) .\label{f}\ee
We thus have, for each double cone $\2O$, a `factorization' of the global symmetry
group into two commuting representations which are localized along half lines.
An immediate consequence of (\ref{f}) and the representation property is
\be U(g)\, U^{\2O}_{L/R} (h)\, U(g)^* = U^{\2O}_{L/R} (ghg^{-1}) ,\label{cov}\ee
which expresses covariance of the disorder operators under global gauge transformations.
Arguing that in view of (\ref{disord}) the operators $U^\2O_L(g),U^\2O_R(g)$ are
associated to the double cone $\2O$ we define the following extension of the field 
algebras:
\be \hat{\2F}(\2O)=\2F(\2O) \vee U^{\2O}_L (G)'' .\ee
Whereas this enlarged net obviously is nonlocal one can still verify isotony
$\2O_1\subset\2O_2 \ \impl \ \hat{\2F}(\2O_1)\subset\hat{\2F}(\2O_2)$
and Poincar\'{e} covariance. Furthermore, the extended net is not too large in
the following sense:
\be \hat{\2F}(\2O)\wedge\2A(\2O)'=\7C\11 \ \ \forall\2O\in\2K.\ee
Due to (\ref{cov}) the algebras $\hat{\2F}(\2O)$ are stable under the extension of
the global group action. This allows to define the fixpoint net
\be \hat\2A(\2O) = \hat{\2F}(\2O) \cap U(G)' ,\label{ahat}\ee
which gives rise to the following square of inclusions for each double cone $\2O$:
\be\ba{ccc} \hat{\2A}(\2O) & \subset & \hat{\2F}(\2O) \\ \cup & & \cup \\
     \2A(\2O) & \subset & \2F(\2O) 
.\ea\label{square}\ee
In this diagram one can go from the right column to the left by restricting to the
invariant elements under $G$. Furthermore, one can show $\hat{\2F}(\2O)$ to be isomorphic
to the crossed product of $\2F(\2O)$ by the automorphism group 
$\alpha^\2O_g=Ad\,U^\2O_L(g)$. In order to simplify the exposition from now on we assume
the group $G$ to be finite. Most of our results remain valid for compact groups, see 
\cite{mue1}. The inclusion $\2A(\2O)\subset\hat{\2F}(\2O)$, being
irreducible and composed of the finite-index depth-2 inclusions 
$\2A(\2O)\subset\2F(\2O)$ and $\2F(\2O)\subset\hat{\2F}(\2O)$, is known to
be of depth 2, too. This amounts to the existence of a finite dimensional Hopf algebra
$H$ acting on $\hat{\2F}(\2O)$ such that $\2A(\2O)$ is the fixpoint algebra 
$\hat{\2F}(\2O)^H$. In the next section this structure will be analyzed quite explicitly.

\section{SPONTANEOUSLY BROKEN QUANTUM DOUBLE SYMMETRY}
We have already remarked that the algebra $\hat{\2F}(\2O)$ is isomorphic to a crossed
product which is equivalent to the existence of a $G$-gradation. This implies that every
$\hat{F}$ has a unique representation of the form
\be \hat{F}=\sum_{g\in G} F(g) \, U^\2O_L(g), \ \ \ F(g)\in\2F(\2O) .\ee
Given an arbitrary function $f\in\7C(G)$ on the group there is thus an action $\gamma_f$:
\be \gamma_f\left(\sum_{g\in G} F(g)\,U^\2O_L(g)\right)= \sum_{g\in G} 
   f(g)\,F(g)\,U^\2O_L(g) ,\ \ F(g)\in\2F(\2O), f\in \7C(G). \label{gamma_F}\ee
In particular, for the delta-functions $\delta_g(h)=\delta_{g,h}$ we obtain the 
projections $\gamma_g:=\gamma_{\delta_g}$. Due to (\ref{cov}) we have
\be \alpha_g\circ\gamma_h=\gamma_{ghg^{-1}}\circ\alpha_g .\label{cov2}\ee

We are now prepared to exhibit the action of the quantum double $D(G)$ on the extended
algebras. Let $\7C(G)$ be the algebra  of (complex valued) functions on the finite
group $G$ and consider the adjoint action of $G$ on $\7C(G)$ according to 
$\alpha_g:\ f\mapsto f\circ Ad(g^{-1})$. The quantum double $D(G)$ is defined 
as the crossed product $D(G)=\7C(G)\rtimes_\alpha G$ of $\7C(G)$ by this
action. In terms of generators, $D(G)$ is the $*$-algebra generated by unitary and
selfadjoint, respectively, elements $U_g,\, V_h,\, g,h\in G$ with the relations
\be U_g\,U_h=U_{gh}, \ \ V_g\,V_h=\delta_{g,h}V_g, \ \ U_g\,V_h=V_{ghg^{-1}}\,U_g  \ee
and the identification $U_e=\sum_g V_g=\11$. 
The action of $D(G)$ is now defined for the basis $\{V_g U_h,\ g,h\in G\}$ by
\be \gamma_{V_g U_h}(\hat{F})=\gamma_g\circ\alpha_h(\hat{F}) \ee
and for $D(G)$ by linear extension. One easily verifies 
$\gamma_{ab}(x) = \gamma_a \circ \gamma_b(x)$ and $\gamma_\11(x)  = x$, whereas
the well-known Hopf algebra maps on $D(G)$ \cite{dpr} lead to
\bea \Delta(V(g)U(h))=\sum_k V(hk)U(h)\otimes V(k^{-1})U(h) &\impl&
    \gamma_a(xy) = \gamma_{a^{(1)}}(x)\gamma_{a^{(2)}}(y) ,\nn\\
  S(V(g)U(h))=V(h^{-1}g^{-1}h)U(h^{-1}) &\impl& (\gamma_a(x))^*=\gamma_{S(a^*)}(x^*),\\
  \ve(V(g)U(h))=\delta_{g,e} &\impl& \gamma_a (\11)=\ve(a)\11 .\nn\eea
(We have used the standard notation $\Delta(a)=a^{(1)}\otimes a^{(2)}$ for
the coproduct.) This proves that $\gamma: D(G)\times\2M\to\2M$ defines an action of 
$D(G)$ on the local algebras $\hat{\2F}(\2O)$. As this action is compatible with the
local structure it extends to a unique action on the quasilocal algebra $\hat{\2F}$.

In the case of an abelian group $G$ this can be reformulated in terms of commuting
actions of $G$ and the dual group $\hat{G}$, the total symmetry group thus being
$G\times\hat{G}$. It is clear that the $\hat{G}$-part of the symmetry is spontaneously
broken in the sense that there are no unitary operators on $\2H$ implementing this
action. The same holds, of course, in the non-abelian case where the symmetry to be
unbroken would mean that there exist operators $U(a)\ \forall a\in D(G)$ such that
\be U(a) \, x = \gamma_{a^{(1)}}(x) \, U(a^{(2)}) .\ee
Despite the partial breakdown of the symmetry one can prove that the spectrum of
the action of $D(G)$ is complete in the sense that for every finite dimensional
representation $D_{ij}$ of $D(G)$ there is a multiplet 
$\psi_i,\ i=1,\ldots,\mbox{dim}(D)$ in each $\hat{\2F}(\2O)$ such that
\be \gamma_a(\psi_i)=\sum_{i'=1}^d D_{i'i}(a) \, \psi_{i'} .\ee
Let now $\psi^1, \psi^2$ be $D(G)$-tensors in $\hat{\2F}(\2O_1),\,\hat{\2F}(\2O_2)$, 
respectively, where $\2O_2$ lies in the left spacelike complement of $\2O_1$. Then one 
can prove the following C-number commutation relations:
\be \psi_i^1\,\psi_j^2 =\sum_{i'j'} \psi^2_{j'}\,\psi^1_{i'}\ 
   (D^1_{i'i}\otimes D^2_{j'j}) (R) ,\label{commrel3}\ee
where $D^1,\,D^2$ are the matrices of the respective representations and 
\be R=\sum_{g\in G} V_g\otimes U_g\in D(G)\otimes D(G) .\ee
The operators $\psi_i$ can be chosen as isometries fulfilling the relations
\be \psi_i^*\,\psi_j = \delta_{i,j}\11,\ \ \ \sum_{i=1}^{d}\psi_i\,\psi_i^* = \11 .\ee
Let $\psi_i,\ i=1\ldots \mbox{dim}(D)$ be a multiplet of isometries in $\hat{\2F}(\2O)$
transforming according to the irreducible representation $D$ of $D(G)$. Then the maps
\be \rho(\cdot) = \sum_i \psi_i \cdot \psi_i^*,\ \ \
   \phi(\cdot) = \frac{1}{\mbox{dim}(D)}\sum_i \psi_i^* \cdot \psi_i \ee
define a unital *-endomorphism of $\hat{\2F}$ and its left inverse \cite{dhr3},
respectively. The 
relative locality of $\2A$ and $\hat{\2F}$ implies the restriction of $\rho$ to $\2A$ 
to be localized in $\2O$ in the sense that $\rho(A)=A\ \forall A\in\2A(\2O')$. 
If the net $\2A$ satisfied Haag duality we could conclude by standard arguments 
\cite{dhr3} that $\rho$ maps $\2A(\2O_1)$ into itself if $\2O_1\supset\2O$. Despite the
fact that Haag duality does not hold for $\2A$, this is still true, however, as 
follows from the $D(G)$-invariance of $\rho(x)$ for $x\in\2A$. We can thus use the 
formalism of \cite{dhr3} to define the statistics operator which can be shown to be
\be \ve(\rho_1,\rho_2)=\sum_{ijkl} \psi_i^{(2)}\,\psi_l^{(1)}\,\psi_j^{(2)*}\,
  \psi_k^{(1)*}\ (D_{lk}^1\otimes D_{ij}^2)(R) .\label{statop}\ee
Computing $\lambda_\rho=\omega_\rho/d_\rho\equiv\phi_\rho(\ve(\rho,\rho))$, the 
statistics dimension $d_\rho$ turns out to coincide with the dimension of the 
representation $D$ of $D(G)$ whereas the statistics phase $\omega_\rho$ is given by
\be \omega_\rho\,\delta_{ij}=D_{ij}(X),\ \ \ X=\sum_g V_g\,U_g\in D(G) ,\ee
where the central unitary element $X\in D(G)$ is just the (inverse of the) `ribbon 
element' \cite{rt} of the modular Hopf algebra $D(G)$. Finally, defining the monodromy 
operators $\ve_M(\rho_1,\rho_2)=\ve(\rho_1,\rho_2)\ve(\rho_2,\rho_1)$ we can compute
the statistics characters \cite{khr}:
\be Y_{ij}\equiv d_i d_j\,\phi_i(\ve_M(\rho_i,\rho_j)^*)=(tr_i\otimes tr_j)\circ
  (D^i\otimes D^j)(I^*), \ee
where $I=R\sigma(R)$ is again well known in the context of modular Hopf algebras.
We have thus established, for a special class of models, a  correspondence 
between the notions of algebraic QFT and those of \cite{rt}. Yet, the framework is
not exactly as in \cite{dhr3,frs}. The point is that one can prove \cite{mue2} that our 
assumptions, in particular the split property for wedges, preclude the existence
of nontrivial DHR sectors. That this no-go theorem does not apply in the present
situation is due to the fact, to be discussed in the rest of this note, that Haag
duality does not hold for the fixpoint net $\2A$.

\section{HAAG DUALITY}
We will now comment on a less well known two-dimensional pecularity, namely the fact 
that the step \cite{dhr1} from (\ref{twduality}) to (\ref{duality}) fails in 1+1 
dimensions. This means that one cannot conclude from (twisted) duality of the fields 
that duality holds for the observables in simple sectors, which in fact is violated.
The origin of this phenomenon is easily understood.
Let $\2O\in\2K$ be a double cone. One can then construct gauge invariant
operators in $\2F(\2O')$ which are obviously contained in $\2A(\2O)'$
but not in $\2A(\2O')$. This is seen remarking that the latter algebra, belonging 
to a disconnected region, is defined to be generated by the observable algebras 
associated to the left and right spacelike complements of $\2O$, respectively.
This algebra does not contain gauge invariant operators constructed using fields
localized in both components. The weaker property of {\it wedge duality} remains
true, however. Let $\2H_1$ be a simple sector. Then:
\be \6A(W)\equiv\2A(W)\restr\2H_1\ \Longrightarrow\ \6A(W)'=\6A(W')\ \   \forall W
\label{w-duality} .\ee
Defining now the {\it dual net} by
\be \6A^d(\2O)=\6A(W^\2O_L)\cap\6A(W^\2O_R) \ee
it is easy to verify that Haag duality holds for $\6A^d$. One would, however, like
to know which additional operators are obtained in this way. Using the above methods
we can actually compute the dual net in terms of $\2A(\2O)$ and the disorder operators.
One can show that the net $\hat{\2A}$ defined in (\ref{ahat}) is local and
leaves the sectors in $\2H$ invariant so that it constitutes a local extension of
$\6A$ in each sector. Using the formula
\be \2A(\2O')' = \2F(\2O)\vee U^\2O_L(G)''\vee U^\2O_R(G)''\ee
one can in fact prove it to coincide with the dual net: 
$\hat{\2A}(\2O)\restr\2H_1=\6A^d(\2O)$ for every simple sector $\2H_1$.
This is reminiscent of the analysis in \cite{rob} where nets of observables (in at 
least 2+1 dimensions) which arise as fixpoints under a group of inner symmetries from
a field theory were shown to violate Haag duality whenever the symmetry is 
spontaneously broken in the sense that the vacuum is not invariant under the whole 
group. Again the observables fulfill a weaker property ({\it essential duality}) which
allows to construct a maximal local extension satisfying Haag duality. 
This dual net was shown in \cite{rob} to be just the fixpoint net of the field net under 
the {\it unbroken} part of the gauge group. The analogy to the situation studied above
is obvious, for here $\hat{\2A}=\hat{\2F}^G$ are the invariants under the unbroken part 
$G\subset D(G)$ of the quantum double.

\section{ACKNOWLEDGEMENTS}    
I am grateful to the organizers of the Carg\`{e}se summer school for the opportunity
to present this seminar. Thanks are also due to K.-H.~Rehren for many useful discussions
and to the Studienstiftung des deutschen Volkes for financial support.

\numbibliography
\bibitem{bern}D. Bernard, A. LeClair: The quantum double in integrable \qft, 
   \npb {\bf 399}(1993)709

\bibitem{bdl}D. Buchholz, S. Doplicher, R. Longo: On Noether's theorem in \qft,
   Ann. Phys. {\bf 170}(1986)1

\bibitem{dvvv}R. Dijkgraaf, C. Vafa, E. Verlinde, H. Verlinde: The operator
   algebra of orbifold models, \cmp {\bf 123}(1989)485

\bibitem{dpr}R. Dijkgraaf, V. Pasquier, P. Roche: Quasi Hopf algebras, group
   cohomology and orbifold models, \npb (Proc. Suppl.){\bf 18B}(1990)60

\bibitem{dhr1}S. Doplicher, R. Haag, J. E. Roberts: Fields, observables and
   gauge transformations I, \cmp {\bf 13}(1969)1

\bibitem{dhr3}S. Doplicher, R. Haag, J. E. Roberts: Local observables and particle
   statistics I+II, \cmp {\bf 23}(1971)199, {\bf 35}(1974)49

\bibitem{dr2}S. Doplicher, J. E. Roberts: Why there is a field algebra with a 
   compact gauge group describing the superselection structure in particle physics, 
   \cmp {\bf 131}(1990)51

\bibitem{drin1}V. G. Drinfel'd, Quantum groups, {\it in:} Proc. Int. Congr. Math.,
   Berkeley 1986

\bibitem{frs}K. Fredenhagen, K.-H. Rehren, B. Schroer: Superselection sectors 
   with braid group statistics and exchange algebras I. General theory, \cmp
   {\bf 125}(1989)201

\bibitem{haag}R. Haag: {\em Local Quantum Physics}, 2nd ed., Springer, 1996

\bibitem{K}D. Kastler (ed.): The algebraic theory of superselection sectors. 
   Introduction and recent results, World Scientific, 1990

\bibitem{mue1}M. M\"uger: Quantum double actions on operator algebras and orbifold
   quantum field theories, preprint DESY 96-117 and hep/th-9606175

\bibitem{mue2}M. M\"uger: The superselection structure of massive \qfts\ in $1+1$
   dimensions, in preparation

\bibitem{khr}K.-H. Rehren: Braid group statistics and their superselection rules,
   {\it in}: \cite{K}

\bibitem{rt}N. Yu. Reshetikhin, V. G. Turaev: Invariants of 3-manifolds via link
   polynomials and quantum groups, Invent. Math. {\bf 103}(1991)547

\bibitem{rob}J. E. Roberts: Spontaneously broken gauge symmetries and 
  superselection rules, in: G. Gallavot\-ti (ed.): Proc. International School of 
  Mathematical Physics, Camerino 1974   

\bibitem{sum}S. J. Summers: On the independence of local algebras in \qft, 
   \rmp {\bf 2}(1990)201

\bibitem{szlvec}K. Szlach\'{a}nyi, P. Vecserny\'{e}s: Quantum symmetry and braid
   group statistics in G-spin models, \cmp {\bf 156}(1993)127

\endnumbibliography
\end{document}